\documentclass[12pt]{article}

 \usepackage{amsfonts}
 \usepackage{amssymb,amsmath}
 \usepackage{graphicx} \usepackage{epstopdf}
 \newcommand{\crlb}[1]{\label{#1}\\[2pt]}
 
 \newcommand{\crld}[1]{\label{#1}}
 \newcommand{\eela}[1]{\quad\hbox{\scriptsize{#1}}\label{#1}\end{eqnarray}}
 \newcommand{\eelb}[1]{\label{#1}\end{eqnarray}}
 
 \newcommand{\newsecb}[2]{\section{#1}\label{#2}\setcounter{equation}{0}}

 \newcommand{\nolabels} {\def\eel{\eelb}\def\eeql{\eeqlb}  \def\crl{\crlb} \def\newsecl{\newsecb}\def\bibiteml{\bibitem} \def\citel{\cite}\def\labell{\crld}}
\newcommand{\eeqla}[1]{\quad\hbox{\scriptsize{#1}}\label{#1}\end{aligned}\end{equation}}
\newcommand{\eeqlb}[1]{\label{#1}\end{aligned}\end{equation}}

\newcommand\publishversion{\nolabels\setlength{\textheight}{8.3in}\setlength{\oddsidemargin}{0in}
   	 \setlength{\textwidth}{6.3in}\setlength{\topmargin}{-0.2in}}
\def\beq{\begin{equation}\begin{aligned}}		\def\eeq{\end{aligned}\end{equation}}
\def\be{\begin{eqnarray}}  					\def\ee{\end{eqnarray}}		%\be and \ee will become obsolete in due time.
\def\bi#1{\begin{itemize}\item[#1]} 			 			\def\ei{\end{itemize}} 
  \def\eqn#1{(\ref{#1})}
   	 \def\fn{\footnote}	  		 

		    \def\b{\beta}   \def\d{\delta}          
		 \def\L{\Lambda}      \def\r{\varrho}          
    		  	   \def\e{\varepsilon} 
	    		        		     		\def\vv{\varphi}     
 	 		\def\s{\sigma}     	\def\tht{\theta}

\def\OO{{\mathcal O}} 		     
 		      
			\def\ra{\rightarrow}	
 		\def\ket{\rangle}

\def\fract#1#2{{\textstyle\frac{#1}{#2}}}	 	 	
\def\ffract#1#2{\raise .2 em\hbox{$\scriptstyle#1$}\kern-.3em/\kern-.2em\lower .15 em \hbox{$\scriptstyle#2$}}
\def\half{\fract12}					
 
\def\ex#1{e^{\textstyle#1}} 				
\def\bpmatrix{\begin{pmatrix}} 			\def\epmatrix{\end{pmatrix}}
\def\bmatrix{\begin{matrix}} 			\def\ematrix{\end{matrix}} 
\def\bcenter{\begin{center}}			\def\ecenter{\end{center}}

\def\lowerheightfig#1#2#3{\(\raise-#1\hbox{\includegraphics[height=#2]{#3}}\)}
\def\lowerwidthfig#1#2#3{\(\raise-#1\hbox{\includegraphics[width=#2]{#3}}\)}
		 
\def\weglaten#1{}		
\def\BH{{\mathrm{BH}}}\def\Pl{{\mathrm{Planck}}}\def\inn{{\mathrm{in}}}
\def\outt{{\mathrm{out}}}
\publishversion 

\begin{document}

\begin{titlepage}
 \title{ \LARGE\bf Discreteness of Black Hole Microstates\\[10pt]}
\author{Gerard 't~Hooft}
\date{\small  Institute for Theoretical Physics \\ Utrecht University  \\[10pt]
 Postbox 80.089 \\ 3508 TB Utrecht, the Netherlands  \\[10pt]
e-mail:  g.thooft@uu.nl \\ internet: 
http:/\!/www.staff.science.uu.nl/\~{}hooft101/ } \maketitle

\noindent \textbf{Abstract}  \begin{quote}
	It is explained that, for black holes much heavier than the Planck mass, black hole microstates can be well understood without string theory. It is essential to understand the antipodal identification at the horizon. We show why the microstates exhibit a discrete spectrum, and how they relate to the particles outside the hole.
 \end{quote} \vfill

\newsecl{The algebra leading to the microstates}{micro}
	Consider a Schwarzschild black hole in a stationary situation, during a time interval \(\OO(M_\BH\log M_\BH)\) in Planck units. Let it be surrounded by particles with masses and energies \(E_i\ll M_\Pl\ll M_\BH\), and densities in the range of that of Hawking particles. There are particles moving inwards and outwards. The black hole mass, \(M_\BH\), fluctuates accordingly when particles are absorbed and emitted. During the given time interval these mass fluctuations are small compared to the entire black hole mass. Consequently, we can handle the black hole metric as a background metric, with minor perturbations described by perturbative (quantum) gravity\,\cite{GtHantipodes}.
	\end{titlepage} 
	
This time-reversal symmetric situation is the best configuration for describing quantum energy eigen states, since these have the same time symmetry. String theory\,\cite{GSW,Polchinski} is claimed to account for these quantum states by appealing to pictures of stacks of \(D\)-branes\,\cite{Dbranes}, fuzzballs\,\cite{fuzzballs} and other concoctions, while it appears to be almost impossible to reconstruct space-time itself, and the description of events as experienced by observers moving in or out, or to understand how unitarity is maintained in the black hole evolution. Where does quantum information go?
	
In recent papers\,\cite{GtHantipodes, GtHrecent} the author observed that, to get the picture right, one has to implement the gravitational back reaction between in- and out-states, and to impose the antipodal identification constraint. Disregarding these important insights inevitably leads to imprecise formulations. In contrast to claims often made, one cannot make the horizon disappear by becoming``fuzzy", or believe that strings already take over at macroscopic distance scales, without serious damage to the concept of General Relativity.
\setcounter{page}{2}

In a locally regular coordinate frame such as the Kruskal-Szekeres frame, the past event horizon is given by \(u^+=0\), and the future event horizon is \(u^-=0\), where \(u^+\) and \(u^-\) are local light cone coordinates. The most important region where most of the non-trivial physics takes place is the region where these two horizons intersect.  This region is often portrayed as an \(S_2\) sphere, but we noticed\,\cite{GtHantipodes} that it has to be replaced by a \emph{projective} sphere, \(S_2/\mathbb Z_2\). So, at the intersection, we identify antipodal points. The fact that the physical region is defined by \(r\ge 2M_\BH\) implies  that \(r\) never comes close to zero. Therefore, the \(\mathbb Z_2\) identification never generates any physical singularity. As for the central singularity of the Schwarzschild black hole, it is well separated from the physical domain(s) by a horizon (cosmic censorship works fine here), so that it can do no harm.

Particles entering the black hole, in-particles for short, may be represented by the momentum distribution, \(p^-(\tht,\,\vv)\) that they deposit on the future event horizon. The out-particles can be characterised by their positions \(u^-(\tht,\,\vv)\) when they leave the past event horizon. A condition for our information preservation process to work is that these operator distributions suffice to characterise all in-states and all out-states. There are good reasons to assume that this should work. In any case, these are the variables in terms of which we shall define all states, both for the particles deeply imbedded in the black hole\fn{that is, close to an event horizon, but at the \emph{physical} side of it.} and for the particles further outside. Outside observers never need to refer to particles `inside' the black hole, that is, across any of the horizons.

The \emph{quantum commutation rules} turn out to be 
\be [u^\mp(\tht,\,\vv),\ p^\pm(\tht',\,\vv')]=i\,\d(\tht-\tht')\,\d(\vv-\vv')\sin^{-1}\tht\ , \eel{commup}
while\fn{Some caution is asked for: time \(t\) is not an operator in the usual sense, so states are characterised \emph{either}
by giving their wave function on the \(u^+\) axis \emph{or} on the \(u^-\) axis.} \( [u^\pm(\tht,\vv),\ u^\pm(\tht',\vv')]=[p^\pm(\tht,\vv),\ p^\pm(\tht',\vv')]=0\)\,. 

To describe the scattering matrix, we expand both \(u^-\) and \(p^-\) in spherical harmonics \(Y_{\ell\,m}(\tht,\vv)\), turning them to variables \(u^\pm_{\ell\, m}\) and \(p^\pm_{\ell\, m}\), obeying
\be  [u^\mp_{\ell\, m},\ p^\pm_{\ell'\,m'}]=i\,\d_{\ell\,\ell'}\,\d_{m\,m'}\ , \eel{commellm}

The condition that the masses and the momenta of the particles are all demanded to stay below the Planck regime ensures that we need not worry about fluctuations of the space-time metric, but this does restrict the time period considered to stay well within the domain \(\OO(M_\BH\log M_\BH)\). The earliest in-particles and the latest out-particles then develop momenta close to or beyond the Planck mass. These we now address by the back reaction equations\,\cite{GtHantipodes}:
	\be u^{-\,(\outt)}_{\ell\,m}=\frac{8\pi G}{\ell^2+\ell+1}\,p^{-\,(\inn)}_{\ell\,m}\ ,\qquad 
	u^{+\,(\inn)}_{\ell\,m}=-\frac{8\pi G}{\ell^2+\ell+1}\,p^{+\,(\outt)}_{\ell\,m}\ .\qquad \eel{backreac}
Consequently, the coordinates \(u^+\) and \(u^-\) do not commute,
	\be { }[u_{\ell\,m}^{+\,(\outt)},\,u_{\ell' m'}^{-\,(\inn)}]=\frac{8\pi\,G\,i}{\ell^2+\ell+1}\,\d_{\ell \ell'}\,\d_{mm'}\ . \eel{uucomm}
The suffixes `(in)' and `(out)', are superfluous and will be omitted henceforth.
The relations \eqn{commellm} --- \eqn{uucomm} do not allow us to restrict ourselves to the domain \(I\), defined by
(\(u^+>0,\ u^-<0\)), since the Fourier transform of a function supported by a domain \(x>0\) cannot be restricted to a domain \(p>0\).
This is why region \(I\) cannot be considered separately from region \(II\), defined by  (\(u^+<0,\ u^->0\)). Since both regions must be physical, we assume that region \(I\) refers to one hemisphere of the black hole only, while region \(II\) refers to the other hemisphere. It is the only way to keep the wave functions pure, while the condition that no cusp singularities are allowed makes this assignment unique. This is the antipodal identification\,\cite{GtHantipodes}.
		
We do see that, if \(|p_{\ell,m}^{-\,(\inn)}|>\ell^{\,2}\,M_\Pl\), then the in-particle responsible for this large value of the momentum can be replaced by the associated out-particle, since its position now is far enough reparated from the point \(u^-=0\) to be considered soft, as  its momentum will always be in the order of the inverse of its position.

\begin{figure}[h]\begin{center}\lowerwidthfig{0pt}{300pt}{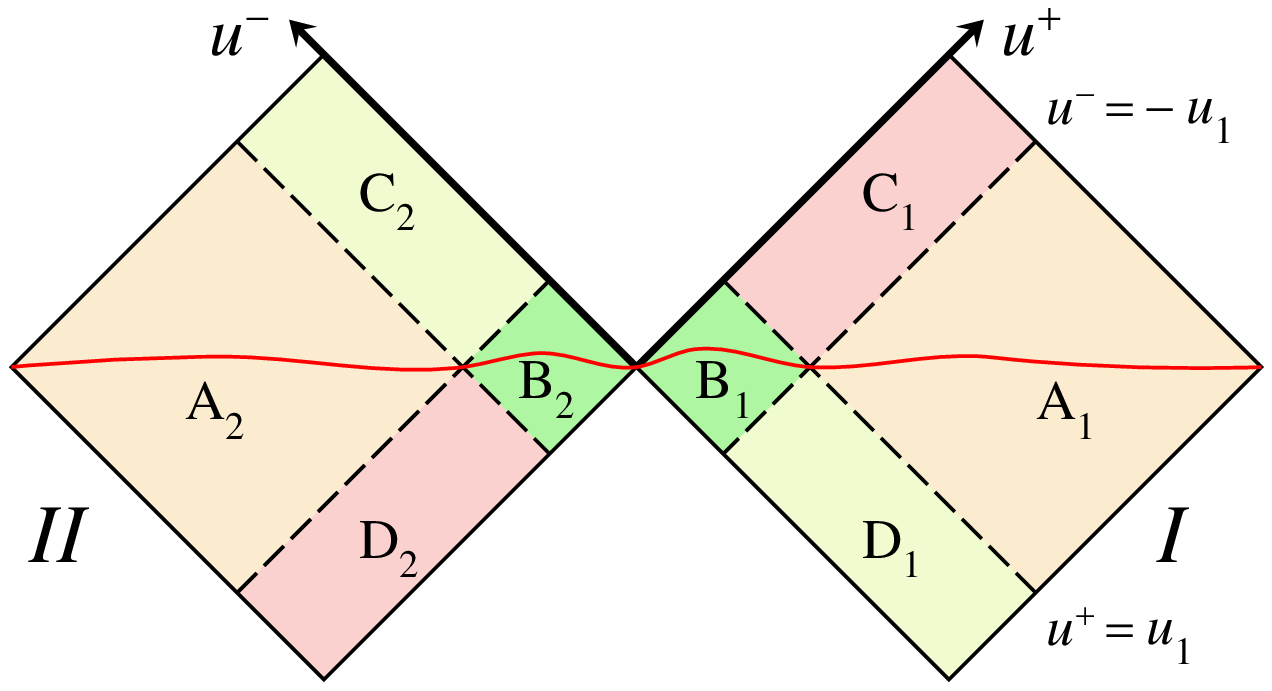} \small
\begin{caption}{\small Dividing regions \(I\) and \(II\) both into four parts}\labell{dividinglines} \end{caption}
to separate the outside of a black hole from the inside.\\ The red line is a Cauchy surface.
\end{center}
\end{figure}

\newsecl{Discreteness}{discr}
Now, in region \(I\), consider two dividing lines: \(u^+=u_1,\ u^-= -u_1\), and similarly region \(II\), see Fig. \ref{dividinglines}.
To count the quantum states, we consider all states  on a Cauchy surface. Choose as our Cauchy surface a line such as the red line in Fig.~\ref{dividinglines}. We see that then the complete set of quantum states is generated by the product of the states in regions  \(A_1,\ B_1,\ B_2\) and \(A_2\). This gives us all available quantum states\fn{Of course, all \emph{entangled} states are just superpositions of these.}. The states in regions \(A_1\) and \(A_2\) represent all particles outside the black hole. They form a continuum, since these domains are not compact. We shall be interested in the states in \(B_1\) and \(B_2\). They may be regarded as states representing the microstates
inside the black hole, although they are still physically within reach for the outside world.

In earlier treatises, these states formed continua also. This is because \(u^+\) and \(u^-\) are tortoise coordinates: considering them as the exponents of regular coordinates, by writing  \(u^\pm=\s^\pm e^{\r^\pm}\), one finds that \(\r^+\) decreases linearly and \(\r^-\) increases linearly in time. Since they are both unbounded below, the quantum states would seem to be continuous. This would have given us far too \emph{many} micro-states, and furthermore, the particles moving inwards seemed to be unrelated to the ones going out (the black hole information paradox).

In our treatise, due to the gravitational back reaction, which solves the information paradox, see Eqs.~\eqn{backreac}, this is different. We have for our microstates:
	\be -u_1<u^-<u_1\ .\eel{compactdomain}
On this domain, the states are spanned by the functions
	\be \sin( \pi n(u+u_1)/2u_1))\ , \ee
having momenta \(|p^+|= \pi  n/(2u_1)\ ,\qquad n>0\ .\) \\
Taking \emph{periodic} boundary conditions gives us a similar set of states with the same density:
	\be p^+=2\pi n^+/2u_1\ , \qquad -\infty<n^+<\infty\ . \ee
But we have the relation \eqn{backreac},
	\(p^+=-\frac{\ell^2+\ell+1}{8\pi G}\,u^+\ ,\)
while also \(u^+\) is bounded,
	\be -u_1<u^+<u_1 \ \ \ra \ \ |p^+|<\frac{(\ell^2+\ell+1)u_1}{8\pi G}\ .	\ee
this implies that, at every \(\ell,\,m\), we have only a finite number of states:
	\be |n^+|=\frac {u_1|p^+|}\pi<\frac{(\ell^2+\ell+1)\,u_1^2}{8\pi^2\,G}\ .
	\ee
If we take the separation line at \(r=2GM(1+\e)\), this brings  the number of states at \(\ell,m\) to
	\be n_\mathrm{max}=\frac{G M^2(\ell^2+\ell+1)\e}{\pi^2}\ \equiv n_0\,(\ell^2+\ell+1)\ .\ee
The total number of states is the product of the number at each \(\ell\). Only odd values of \(\ell\) are allowed. Thus we get:	\def\maxx{{\mathrm{max}}}
	\be N_\mathrm{total}=\prod_{|m|\le\ell<\ell_\maxx}n_\maxx(\ell) = \ex{\, \half\ell^{\,2}_\maxx\log( n_0\,	\ell_\maxx^{\,2}	)
	\big(1+\OO(1/\ell_\maxx)\big) } \ .	\ee
At this point, there is one difficulty left: how many values of \(\ell\) and \(m\) should we admit? It seems reasonable to impose\fn{This is needed if we want not only discreteness, but also a strict maximum of the number of microstates. We are still working on finding a more precise treatment to relate this maximum to the area of the horizon.}  a maximal value \(\ell_\maxx\) for \(\ell\). In that case, we find that Hawking's value for the total entropy would match if, in Planck units,
	\be \ell_\maxx^2\log\ell_\maxx=\OO(M_\BH^2)\ ,\ee
which is the domain of values where the angular momenta of the in- and out-particles is near the maximal value for capture by the black hole.

Note, that in general, Hawking radiation is dominated by the lowest \(\ell\) values; in the wave functions, higher \(\ell\) is strongly suppressed.

	\newsecl{Discussion}{discn}
There is an important caveat. In Eq.~\eqn{compactdomain}, it was assumed that large values for \(u_1\) would imply that particles are far separated from the horizon. We must be aware that we are actually discussing the \((\ell,\,m)\) component of the partial wave expansion. This expansion describes the positions as functions of \(\tht\) and \(\vv\). In reality we should only attach one value set for \(\tht\) and \(\vv\), not one value set for \(\ell\) and \(m\). Therefore, we should regard the above derivations as a first approach to the discreteness of the microstates, but this is not the final word.

It would be more elegant if we could make the following train of arguments more rigorous: the Hartle-Hawking wave function is the entangled state
	\be |\psi\ket_{HH}=C\sum_{E,n}e^{-\half\b E}|E,n\ket_I\,|E,n\ket_{II}\ , \eel{HH}
where \(\b\) is the inverse Hawking temperature, \(E\) the energies of the states both in region \(I\) and region \(II\), and \(n\) stands short for any other type of quantum numbers. For a local observer near the horizon, \(|\psi\ket_{HH}\)  represents the single vacuum state, while for the distant observer it contains Hawking particles both going in and out. Averaging over the unseen states in region \(II\) gives us the thermal mixed states associated to Hawking's temperature. His value for the entropy, as the logarithm of the number of microstates, follows directly. In our picture the precise interpretation is slightly different. When we are close to equilibrium, the two hemispheres of the black hole are entangled.
 An observer watching at most one hemisphere of the black hole (regardless in which orientation) will not notice the entanglement, and hence observe the same thermally mixed state. Now if the local observer sees a vacuum, the global observer actually sees the entangled state \eqn{HH}.  The second hemisphere is entangled with the first, and hence the total number of microstates reached is strictly speaking only one, in practice better described as a combination of the states \(|E_1,n_1\ket_I\,|E_2,n_2\ket_{II}\), which we also get when things are thrown into the hole. Far from equilibrium, the local observer will see particles moving in and out, different everywhere; the external observer should then see all the states.
 
 In the present paper, our only concern was to establish that the microstates should have a discrete spectrum; this we think we have demonstrated.
 There is the question of the cut-off \(\ell_\maxx\) at high \(\ell\), which requires further exploration. At high \(\ell\) the gravitational back reaction produces effects in the transverse directions, and the different \(\ell,m\) partial waves begin to interact with one another.
 
 The non-gravitational interactions were ignored. In general quantum field theories, the couplings are weak, and they stay weak at the horizon, unlike the gravitational couplings. There are ways however to improve the argument, for instance by including gauge theory contributions. We do not expect these to substantially affect our conclusions.

Needless to state that string theories and AdS/CFT conjectures  were bypassed in our analysis. We are in 3+1 space-time dimensions, and have flat asymptotic space-time (\(\L=0\)). There is no need for supersymmetry, and there is no need to go to the BCS limit, where horizons are quite different from the more representative Schwarzschild case (in the BCS limit, the Cauchy horizon and the event horizon almost coincide, to form a structure that is quite different from an ordinary event horizon). No exotic assumptions had to be made to understand the gravitational back reaction, and, although the antipodal identification is an assumption, it is a natural assumption concerning space-time topology that does not violate causality and actually restores unitarity for the black hole. It is the only way to avoid cusp singularities at the horizon.

As we only considered black holes close to equilibrium, the question how antipodal identification switches on in the black hole formation process was not answered, but we may assume that this will be genuine Planckian physics that is not yet understood. Note that, when a black hole forms, the horizon starts out as almost a single point, where `antipodal identification' would only span Planckian distance scales.

\end{document}